\documentclass[11pt,notitlepage,prd,a4paper,
preprintnumbers,superscriptaddress,nofootinbib]{revtex4-1}

\usepackage[utf8]{inputenc}
\usepackage[colorlinks=true,citecolor=blue,linkcolor=blue]{hyperref}
\usepackage[normalem]{ulem}
\usepackage{url}
\usepackage{graphicx,wrapfig,float,slashed,cancel}
\usepackage{amsmath,amssymb,epsfig,graphicx,xcolor,stmaryrd}
\usepackage{bm}
\usepackage{enumitem}
\usepackage{multirow}

\definecolor{darkblue}{RGB}{1, 90, 173}

\def\nnb{\nonumber}

\begin{document}


\title{Light cone QCD sum rules study of the rare radiative 
  $\Xi^{*}_{bb}\to\Xi_b\gamma$ decay}

\author{T.~M.~Aliev}
\email{ taliev@metu.edu.tr}
\affiliation{Physics Department, Middle East Technical University, 06531, Ankara, Turkey}
\author{A.~Ozpineci}
\email{ozpineci@metu.edu.tr}
\affiliation{Physics Department, Middle East Technical University, 06531, Ankara, Turkey}
\author{Y.~Sarac}
\email{ yasemin.sarac@atilim.edu.tr}
\affiliation{Electrical and Electronics Engineering Department,
Atilim University, 06836 Ankara, Turkey}

\date{\today}

\preprint{}

\begin{abstract}

The rare radiative decay $\Xi^{*}_{bb}\to\Xi_b\gamma$ is investigated within the light cone QCD sum rules approach. This decay proceeds through the flavor changing neutral current  $b\to s$ transition in the Standard Model. The hadronic matrix element of the considered decay is parameterized in terms of four tensor form factors $T_1^{V}(q^2)$, $T_2^{V}(q^2)$, $T_1^{A}(q^2)$ and $T_2^{A}(q^2)$. The sum rules for these  form factors describing the $\Xi^{*}_{bb}\to\Xi_b\gamma$ decay are derived at $q^2=0$ point using the $\Xi_b$ distribution amplitudes. The results of the form factors are employed to calculate the corresponding decay width. Our finding indicates that this weak radiative decay could be within reach of future high statistics studies of doubly heavy baryons at LHCb and upcoming facilities.

\end{abstract}


\maketitle

\renewcommand{\thefootnote}{\#\arabic{footnote}}
\setcounter{footnote}{0}
\section{\label{sec:level1}Introduction}\label{intro}

The existence of doubly heavy baryons composed of two heavy quarks \(Q\) and one light quark \(q\) is predicted by the quark model. Doubly heavy baryons are divided into two families, \(\Xi_{QQ}\) and \(\Omega_{QQ}\). The \(\Xi_{QQ}\) family contains a light \(u\) or \(d\) quark, whereas the \(\Omega_{QQ}\) family contains a strange quark \(s\). Experimentally, the first evidence for a doubly heavy baryon was reported by the SELEX Collaboration~\cite{SELEX:2002prl}. In 2017, the LHCb Collaboration observed the doubly heavy baryon \(\Xi_{cc}^{++}\) in the decay channel \(\Xi_{cc}^{++} \to \Lambda_c^{+} K^{-} \pi^{+} \pi^{+}\)~\cite{LHCb:2017prl_observation} with mass \(3621.55 \pm 0.23 (stat) \pm 0.30 (syst)~\mathrm{MeV/c^2}\)~\cite{LHCb:2020jhep_mass}. Later, this state was confirmed in the analyses of  \(\Xi_{cc}^{++} \to \Xi_c^{+}\pi^{+}\) decay~\cite{LHCb:2018prl_decay}. It should be noted that the spin-parity quantum numbers \(J^{P}\) of the \(\Xi_{cc}^{++}\) have not yet been determined. Despite intensive experimental studies, other members of the doubly heavy baryon multiplets have not yet been observed by LHCb and BELLE collaborations. Experimental searches for these states are ongoing within these collaborations~\cite{LHCb:2018prl_lifetime}. These observations have stimulated extensive theoretical studies of the properties of doubly heavy baryons.

The mass spectrum of doubly heavy baryons has been investigated within various theoretical approaches, such as the relativistic quark model \cite{Ebert:2002PRD,Roberts:2008IJMPA,Karliner:2014PRD,He:2004PRD}, QCD sum rules \cite{Bagan:1993PLB,Wang:2010EPJA,Kiselev:2002PU,Zhang:2008PRD,Aliev:2012NPA}, potential models \cite{Richard:2005Bled}, and lattice QCD calculations \cite{Lewis:2001PRD,Flynn:2003JHEP,Liu:2010PRD}, etc. Studies of the electromagnetic and strong decays also play an exceptional role in understanding the dynamics of doubly heavy baryons (see the
review \cite{Chen:2017RPP} and references therein, as well as~\cite{Cheng:2021qpd,Ebert:2004ck,Roberts:2009,Branz:2010pq,Albertus:2010hi,Bahtiyar:2018vub,Qin:2021tvk,Eakins:2012,Xiao:2017udy,Xiao:2017dly,He:2021qbs,Mehen:2017nrh,Ma:2017nik,Yan:2018zdi,Chen:2022prd}).

Flavor–changing neutral current (FCNC) processes governed by the transitions \(b\to s\) constitute a promising area for precise tests of the Standard Model (SM). In the SM this transition is forbidden at tree level and occur only at one–loop level. Hence, such decays allow us to scrutinize the gauge structure of the SM and provide a platform to search for new physics beyond the SM. The semileptonic decays of baryons induced at the quark level by \(b\to s\) transitions represent an important opportunity to obtain more precise information on the properties of doubly heavy baryons. Moreover, weak radiative decays of doubly heavy baryons are highly informative for acquiring additional insight into the dynamics of these
systems.

The present work is devoted to the study of weak radiative decay of spin-$\frac{3}{2}$ \(\Xi^{*}_{bb}\)  baryon within the light-cone QCD sum rules (LCSR) method~\cite{Braun:1997kw}. Note that in LCSR the operator product expansion (OPE) is organized in terms of the \emph{twist} of operators, rather than their dimension as in traditional SVZ-type sum rules~\cite{Shifman:1978bx}.

The paper is arranged as follows. In Sec.~\ref{sec:LCSR} we derive the light-cone QCD sum rules for the transition form factors, at $q^2=0$ point, describing the \(\Xi^{*}_{bb}\!\to \Xi_{b}\gamma\) decay. In our calculations we employ the heavy $\Xi_b$ baryon distribution amplitudes (DAs). In Sec.~\ref{III} we perform the numerical analysis for the form factors obtained in the previous section at \(q^{2}=0\) point. We also provide a numerical prediction for the corresponding decay width. Last section  contains a discussion of the obtained results and concluding remarks.

\section{LCSR for the $\Xi_{bb}^{*}\!\to \Xi_{b}\gamma$  decay form factors}\label{sec:LCSR}

The exclusive decay \(\Xi^{*}_{bb}\!\to \Xi_{b}\gamma\) at the quark level is governed by the
flavor–changing neutral current transition \(b\to s\gamma\)  in the Standard Model (SM). This transition takes place at loop level and is described by an effective Hamiltonian. Within this framework, the exclusive decays are represented as a convolution of short and long distance contributions. The short–distance dynamics are encoded in the Wilson coefficients, while the long–distance parts are described by the matrix elements between the initial and final baryon states, which are parametrized in terms of a set of transition form factors. The form factors belong to the nonperturbative domain of QCD; hence their calculation requires some nonperturbative methods. Among these nonperturbative approaches, the QCD sum rules method occupies a special place because it is based on the fundamental QCD Lagrangian. In the present work we employ the light cone QCD sum rule (LCSR) approach to compute the
relevant form factors for the \(\Xi^{*}_{bb}\!\to \Xi_{b}\gamma\) transition responsible for rare radiative decays.

The rare \(b\to s\gamma\) transition is described by the effective Hamiltonian
\begin{equation}
\mathcal{H}_{\rm eff}
 = -\frac{G_F}{\sqrt{2}}\,V_{tb}V_{ts}^{*}
   \Bigg[\sum_{i=1}^{6} C_i(\mu)\,\mathcal{O}_i(\mu)
         + C_7(\mu)\,\mathcal{O}_7(\mu)
         + C_8(\mu)\,\mathcal{O}_8(\mu)
   \Bigg],
\label{eq:Heff-general}
\end{equation}
where \(C_i(\mu)\) are the Wilson coefficients and \(\mathcal{O}_i(\mu)\) are the local
operators. Explicit expressions for the operators and the Wilson coefficients can be
found in~\cite{Buras:1998lectures}. For \(b\to s\gamma\) the main contribution comes from the operator
\begin{equation}
\mathcal{O}_7
 = \frac{e}{4\pi^{2}}\,
   \bar s\,\sigma_{\mu\nu}\,(m_b R + m_s L)\, b\; F^{\mu\nu},
\qquad
R=\frac{1+\gamma_{5}}{2},\quad L=\frac{1-\gamma_{5}}{2},
\end{equation}
so that the effective Hamiltonian relevant for \(b\to s\gamma\) takes the form
\begin{equation}
\mathcal{H}_{b\to s\gamma}
 = -\,\frac{G_{F}\,e}{4\pi^{2}\sqrt{2}}\,
     V_{tb}V_{ts}^{*}\, C_{7}(m_b)\;
     \bar s\,\sigma_{\mu\nu}\,(m_b R + m_s L)\, b\; F^{\mu\nu}.
\label{eq:Heff-bqg}
\end{equation}

The matrix element of the exclusive
\(\Xi^{*}_{bb}\!\to \Xi_{Q}\gamma\) decay is obtained by sandwiching this effective Hamiltonian
between the initial and final states:
\begin{equation}
\mathcal{M}
 = \big\langle \Xi^{*}_{bb}(p',s')\gamma(q,\varepsilon)\,\big|\,
   \mathcal{H}_{b\to s\gamma}\,\big|\,\Xi_{b}(p,s)\big\rangle .
\end{equation}
Replacing \(\sigma_{\mu\nu} F^{\mu\nu} = -2 i\, \sigma_{\mu\nu}\, \varepsilon^{\mu}\, q^{\nu}\)
in the \(\mathcal{O}_7\) operator, the matrix element of interest can be written as
\begin{equation}
\langle \Xi^{*}_{bb}(p') |\, \bar{s}\, i\sigma_{\mu\nu} q^{\nu}(1+\gamma_{5}) b \,| \Xi_{b}(p) \rangle .
\label{eq:matrix3}
\end{equation}
In general this matrix element, in terms of form factors, is parameterized as in the following way
\begin{eqnarray}
\langle \Xi^{*}_{bb}(p') \,|\, \bar{s}\, i\sigma_{\mu\nu} q^{\nu}(1+\gamma_{5})\, b \,|\, \Xi_{b}(p) \rangle
&=& \bar{u}_{\Xi^{*}_{bb},\rho}(p') \gamma_{5}\Big[
      T_{1}^{V}\,\mathcal{P}_{1}^{\rho\mu}
    + T_{2}^{V}\,\mathcal{P}_{2}^{\rho\mu}
    + T_{3}^{V}\,\mathcal{P}_{3}^{\rho\mu}
\Big] u_{\Xi_{b}}(p) \nonumber\\
&& {} +\; \bar{u}_{\Xi^{*}_{bb},\rho}(p') \Big[
      T_{1}^{A}\,\mathcal{P}_{1}^{\rho\mu}
    + T_{2}^{A}\,\mathcal{P}_{2}^{\rho\mu}
    + T_{3}^{A}\,\mathcal{P}_{3}^{\rho\mu}
\Big]\, u_{\Xi_{b}}(p) .
\label{eq:matrix4}
\end{eqnarray}
The Lorentz structures \(\mathcal{P}_{i}^{\mu}\) are defined as
\begin{eqnarray}
\mathcal{P}_{1}^{\rho\mu} &=& \frac{1}{m^2_{\Xi_{b}}}[q^{\rho}p'^{\mu}-p'.q\,g^{\rho\mu}], \nonumber\\
\mathcal{P}_{2}^{\rho\mu} &=& \frac{1}{m_{\Xi_{b}}}[q^{\rho}\gamma^{\mu}-\slashed{q}\,g^{\rho\mu}], \nonumber\\
\mathcal{P}_{3}^{\rho\mu} &=& \frac{1}{m^2_{\Xi_{b}}}\,[q^2\,g^{\rho\mu}\,- q^{\rho}\,q^{\mu}].
\label{eq:P-basis}
\end{eqnarray}
For real photons, \(\mathcal{P}_{3}^{\rho\mu}\) does not contribute and the following endpoint relations take place at $q^2=0$ point (see also~\cite{Hiller:2021Endpoint}):
\begin{equation}
T_{1}^{V}(0) = T_{1}^{A}(0), \qquad
T_{2}^{V(A)}(0) = T_{2}^{A(V)}(0) \mp \frac{m_{\Xi^{*}_{bb}}}{m_{\Xi_{b}}}\, T_{1}^{A(V)}(0),
\label{eq:endpoint}
\end{equation}
where \(m_{\Xi^{*}_{bb}}\) and \(m_{\Xi_{b}}\) denote the masses of \(\Xi_{bb}^{*}\) and \(\Xi_{b}\) baryons, respectively and $-(+)$ sign holds for $T_{2}^V(0)$ ( $T_{2}^A(0)$) form factor. From the end–point relations it follows that, to describe the
\(\Xi^{*}_{bb}\!\to \Xi_{b}\gamma\) decay at \(q^{2}=0\), it suffices to know only two
form factors, e.g. \(T_{1}^{V}(0)\) and \(T_{2}^{V}(0)\).

For calculation of these form factors we apply light cone QCD sum rules method. We start by considering the following correlation function:
\begin{eqnarray}
\Pi_{\mu\nu}(p,p') &=& i \int d^{4}x\, e^{iq\cdot x}\;
\big\langle 0\big|\, T\!\left\{ J_{\mu}(0)
\,\bar b(x)\, i\sigma_{\nu\alpha} q^{\alpha}(1+\gamma_{5})\, s(x)\;
 \right\} \big| \Xi_{b}(p) \big\rangle ,
\label{eq:corr-7}
\end{eqnarray}
where $J_{\mu}$ is the interpolating current of the doubly–heavy spin–\(\tfrac32\) $\Xi_{bb}^{*}$ baryon, and
\(J_{\nu}^{\mathtt{tr}}=\bar b\, i\sigma_{\nu\alpha} q^{\alpha}(1+\gamma_{5}) s\) is the weak transition current.
The explicit current of the doubly–heavy spin–\(\tfrac32\) baryon is
\begin{eqnarray}
J_{\mu}\ &=& \frac{1}{\sqrt{3}}\,\epsilon_{abc}\,
\Big\{
2\big(q^{a,T} C\,\gamma_{\mu}\, b^{b}\big)\, b^{c}
+\big(b^{a,T} C\,\gamma_{\mu}\, b^{b}\big)\, q^{c}
\Big\},
\label{eq:J-A} 
\end{eqnarray}
where \(C\) is the charge–conjugation matrix, \(\epsilon_{abc}\) is the totally antisymmetric tensor in color space.

According to the standard procedure of the QCD sum rules method, the correlation function is
represented in terms of the form factors of the hadron and calculated via the light cone operator product
expansion (OPE). Sum rules for the relevant physical quantities are obtained by
matching the hadronic representation to the OPE result using the quark–hadron
duality ansatz. The hadronic dispersion relation for correlation function is obtained by calculating the
imaginary part with respect to variable \(p'^2\). Inserting a complete set of baryon
states carrying the same quantum numbers as the interpolating current the following relation is obtained
\begin{eqnarray}
\mathrm{Im}_{\,p'^2}\,\Pi_{\mu\nu}(p,p')
&=& \pi\,\delta\!\left(p'^2 - M_{\Xi^{*}_{bb}}^{2}\right)
\sum_{s'} \big\langle 0 \big| J_{\mu}(0) \big| \Xi^{*}_{bb}(p',s') \big\rangle \nonumber\\
&& \times\;
\big\langle \Xi^{*}_{bb}(p',s') \big|\,
\bar s\, i\sigma_{\nu\alpha} q^{\alpha}(1+\gamma_{5})\, b \,\big| \Xi_{b}(p,s) \big\rangle
\;+\; \cdots.
\label{eq:disp-6}
\end{eqnarray}
Here the sum $\sum_{s'}$ indicates the summation over the polarizations of the intermediate \(\Xi^{*}_{bb}\) baryon; the dots denote higher states and continuum. The matrix element $\langle 0|J_{\rho}(0)|\Xi^{*}_{bb}(p')\rangle$ is defined as
\begin{eqnarray}
\langle 0|J_{\mu}(0)|\Xi^{*}_{bb}(p')\rangle
&=& \lambda\,u_{\mu}(p') ,
\label{eq:coupling-10}
\end{eqnarray}
where $\lambda$ is the residue and $u_{\mu}(p')$ is Rarita-Schwinger spinor for spin-$\frac{3}{2}$ doubly heavy baryon, respectively.

The matrix element of the weak transition current between the initial and final
baryon is determined via the form factors as given in Eq.~(\ref{eq:matrix4}). Using Eqs.~(\ref{eq:matrix4}) and (\ref{eq:coupling-10}) and performing the summation over the spin of the
$\Xi^{*}_{bb}$ baryon as,
\begin{eqnarray}
\sum_{s'} u_{\mu}(p',s')u_{\nu}(p',s')=-(\slashed{p'}+m_{\Xi^{*}_{bb}})\Big[g_{\mu\nu}-\frac{1}{3}\gamma_{\mu}\gamma_{\nu}-\frac{2}{3}\frac{p'_{\mu}p'_{\nu}}{m_{\Xi^{*}_{bb}}^2}+\frac{1}{3}\frac{p'_{\mu}\gamma_\nu-p'_{\nu}\gamma_\mu}{m_{\Xi^{*}_{bb}}}\Big],
\end{eqnarray}
for the hadronic part of the correlation function we get
%
\begin{eqnarray}
\Pi_{\mu\nu}^{\text{(phys)}} 
&=& 
\frac{
   \langle 0 | J_{\mu} | \Xi^{*}_{bb}(p') \rangle\,
   \langle \Xi^{*}_{bb}(p') | \bar{b}\, i\sigma_{\nu\alpha} q^{\alpha}(1+\gamma_{5}) s | \Xi_{b}(p) \rangle
}{
   m_{\Xi^{*}_{bb}}^{2} - p'^{2}
}+\cdots
\nonumber\\
&=& 
\frac{\lambda\,}{m^{2}_{\Xi^{*}_{bb}} - p'^{2}}\,
(\slashed{p'}+m_{\Xi^{*}_{bb}})\!
\left\{
 \frac{1}{m^{2}_{\Xi_{b}}}
 \big(q_{\mu}p'_{\nu} - g_{\mu\nu}\, q\!\cdot\!p'\big)
   \big(T_{1}^{V}\gamma_{5} + T_{1}^{A}\big)\right. \nonumber\\
& +&\left. \frac{1}{m_{\Xi_{b}}}
 \big(q_{\mu}\gamma_{\nu} - g_{\mu\nu}\slashed{q}\big)
   \big(-T_{2}^{V}\gamma_{5} + T_{2}^{A}\big)
\right\}
u_{\Xi_{b}}(p) +\cdots,
\label{eq:Pi-phys}
\end{eqnarray}
where $\cdots$ represents the terms taking contributions from higher states, and the states with spin-$\frac{1}{2}$. Here we would like to make following remarks. The interpolating current $J_{\mu}$ couples not only to spin-$\frac{3}{2}$ state but also spin-$\frac{1}{2}$ state through the matrix element between vacuum and one particle state
\begin{eqnarray}
   \langle 0 | J_{\mu} | B(p',s=\frac{1}{2}) \rangle = (Ap'_{\mu}+ B\gamma_{\mu})u(p',s). 
\end{eqnarray}
This indicates that the terms proportional to the $p'_{\mu}$ or $\gamma_{\mu}$ on the left carries contribution from spin-$\frac{1}{2}$ states also. Therefore, to isolate the sole contribution of spin-$\frac{3}{2}$ state we choose the structures not having these factors. Besides, we consider certain order for Dirac structures, such as $\gamma_{\mu}\slashed{q}\gamma_{\nu}\slashed{p}$, to eliminate the present dependent structures. Finally, after the Borel transformation, the correlation function in physical side takes the form
\begin{eqnarray}
\tilde{\Pi}_{\mu\nu}^{\text{(phys)}} =\tilde{\Pi}_1 \slashed{q}\gamma_5 g_{\mu\nu} + \tilde{\Pi}_2 \slashed{q}\gamma_5 q_{\mu}v_{\nu}+\tilde{\Pi}_3 \slashed{q} q_{\mu}v_{\nu}+ \tilde{\Pi}_4 q_{\mu}v_{\nu}+\cdots,
\label{Eq:Physside1}
\end{eqnarray}     
where $\tilde{\Pi}$ represents Borel transformed correlation function, $v_{\nu}=p_{\nu}/m_{\Xi_{b}}$ and
\begin{eqnarray}
\tilde{\Pi}_1 &=&
\,e^{-\frac{m_{\Xi^{*}_{bb}}^{2}}{M^2}}\,
\frac{\lambda\,(m_{\Xi_{b}}+m_{\Xi^{*}_{bb}})\,
\bigl(m_{\Xi^{*}_{bb}}\,T_{1}^{V}-m_{\Xi_{b}}(T_{1}^{V}+2T_{2}^{V})\bigr)}
{2\,m_{\Xi_{b}}^{2}}, \nonumber\\
\tilde{\Pi}_2 &=&
e^{-\frac{m_{\Xi^{*}_{bb}}^{2}}{M^2}}\,
\frac{\lambda\,T_{1}^{V}}{m_{\Xi_{b}}}, \nonumber\\
\tilde{\Pi}_3 &=&
e^{-\frac{m_{\Xi^{*}_{bb}}^{2}}{M^2}}\,
\frac{\lambda\,T_{1}^{A}}{m_{\Xi_{b}}}, \nonumber\\
\tilde{\Pi}_4 &=&
-\,e^{-\frac{m_{\Xi^{*}_{bb}}^{2}}{M^2}}\,
\frac{\lambda\,\bigl(m_{\Xi^{*}_{bb}}\,T_{1}^{A}+m_{\Xi_{b}}(T_{1}^{A}+2T_{2}^{A})\bigr)}
{m_{\Xi_{b}}}.
\label{Eq:Physside}
\end{eqnarray}
The QCD side of the calculation is performed in deep Euclidean region where $p'^2 \ll 0$. The same correlation function, Eq.~(\ref{eq:corr-7}), is used with the interpolating current given in Eq.~(\ref{eq:J-A}) and weak transition current.  After the application of Wick theorem, the correlation function is expressed in terms of a heavy quark propagator and the distribution amplitudes (DAs) of the heavy $\Xi_b$ baryon as
\begin{eqnarray}
\Pi_{\mu\nu}^{\mathrm{QCD}}(p',p) & = & \frac{2i}{\sqrt{3}}\int\,d^4x e^{iq \cdot x } q^{\nu'}\sum_{i=1}^4\Lambda_i\Big\{ \mathrm{Tr}[\Gamma_i(\sigma_{\nu\nu'}\,(1+\gamma_5))^T \, S_b^T(-x) \, C \, \gamma_{\mu}] \nonumber \\
 &+& S_b(-x)\, \sigma_{\nu\nu'}\,(1+\gamma_5)\,\Gamma_i^T\, C\,\gamma_{\mu} + \Gamma_i\,\sigma_{\nu\nu'}\,(1+\gamma_5) \, S_b^T(-x) \, C \, \gamma_{\mu}] \Big\}\, u_{\Xi_b}(p,s),
 \label{Eq:QCDside1}
\end{eqnarray}
where $S_b(-x)$ is the propagator of the $b$-quark. Eq.~(\ref{Eq:QCDside1}) is calculated  using the DAs for the heavy $\Xi_b$ baryon obtained using heavy quark effective field theory~\cite{Ali:2012zza} and $\Lambda_i$ and $\Gamma_i$ are
\begin{equation}
  \label{eq:DA}
  \begin{aligned}
    \Lambda_1 &= \frac{1}{8}f^{(2)} \psi_2(t_1,t_2)                 & \Gamma_1 &= \bar{\slashed{n}}\gamma_5 C^{-1}, \\
    \Lambda_2 &= -\frac{1}{8} f^{(1)} \psi_{3\sigma}(t_1,t_2)       & \Gamma_2 &= i \sigma_{\xi\varphi} \bar{n}^{\xi} n^{\varphi}\gamma_5 C^{-1}, \\
    \Lambda_3 &= \frac{1}{4} f^{(1)} \psi_{3s}(t_1,t_2)             & \Gamma_3 &= \gamma_5C^{-1}, \\
    \Lambda_4 &= \frac{1}{8}f^{(2)} \psi_{4}(t_1,t_2)               & \Gamma_4 &= \slashed{n}\gamma_5 C^{-1},
  \end{aligned}
\end{equation}
where $\psi_2$, $\psi_{3 \sigma}(\psi_{3s} )$, and $\psi_4$ are the DA's with twist 2, 3, and 4, respectively and the light cone vectors $n$ and $\bar{n}$ are defined as:
\begin{equation}
  \label{eq:n}
  \begin{split}
    n_{\alpha} &= \frac{1}{v x} x_{\alpha} \\
    \bar{n}_{\alpha} &= 2 v_{\alpha} - \frac{1}{v x} x_{\alpha}.
  \end{split}
\end{equation}
The DA's of heavy baryon $\Xi_b$ are defined as
\begin{equation}
  \label{eq:DA}
  \psi(t_1,t_2)=\int_0^\infty dw w\int_0^1 du e^{-iw (t_1 u+t_2 \bar{u})}\psi(u,w),
\end{equation}
with $\bar{u}=1-u$, $t_1=0$, $t_2=vx$  and $w$ is momentum of the light diquark. Considering the coefficients of the structures $\slashed{q}\gamma_5 g_{\mu\nu} $, $\slashed{q}\gamma_5 q_{\mu}v_{\nu}$, $\slashed{q} q_{\mu}v_{\nu}$ and $q_{\mu}v_{\nu}$ we obtain the sum rules for the form factor by matching the results obtained from the QCD side to that of physical side. 
\begin{eqnarray}
\Pi_{\mu\nu}^{\mathrm{QCD}}(p',p)&=& \int \, du \int\, dw\, \Bigg\{
\Big[
-\frac{2}{\sqrt{3}\,\Delta}\Big(
2\,f_{1}\,\hat{\psi}_{3\sigma}(u,w)\,\bar u
-f_{2}\,\psi_{2}(u,w)\,w\,(2m_{b}+\bar u\,w)\nonumber\\
&-& f_{1}\,\psi_{3s}(u,w)\,w\,(m_{b}+2\,\bar u\,w)
\Big)
-\frac{2\,\bar u}{\sqrt{3}\,\Delta^{2}}\Big(
f_{2}\big(\hat{\psi}_{2}(u,w)-\hat{\psi}_{4}(u,w)\big)\,\bar u\,w\,(2m_{b}+\bar u\,w)
\nonumber\\
&+& 2f_{1}\,\hat{\psi}_{3\sigma}(u,w)\,(m_{b}+2\bar u\,w)\,q\!\cdot\! v
\Big)
\,\Big]
\slashed{q}\,\gamma_{5}\,g_{\mu\nu}\, 
-\Big[\frac{4}{\sqrt{3}\,\Delta\,m_{\Xi_{b}}}\,f_{2}\,\psi_{2}(u,w)\,\bar u\,w^{2}\,\nonumber\\
&-&\frac{4}{\sqrt{3}\,\Delta^{2}\,m_{\Xi_{b}}}\,f_{2}\big(\hat{\psi}_{2}(u,w)-\hat{\psi}_{4}(u,w)\big)\,\bar u^{3}\,w^{2}
\Big]
\slashed{q}\,\gamma_{5}\,q_{\mu}\,v_{\nu}\, 
-\Big[\frac{4}{\sqrt{3}\,\Delta\,m_{\Xi_{b}}}\,f_{2}\,\psi_{2}(u,w)\,\bar u\,w^{2}\nonumber\\
&-&\frac{4}{\sqrt{3}\,\Delta^{2}\,m_{\Xi_{b}}}\,f_{2}\big(\hat{\psi}_{2}(u,w)-\hat{\psi}_{4}(u,w)\big)\,\bar u^{3}\,w^{2}
\Big]\,
\slashed{q}\,q_{\mu}\,v_{\nu}\,
+\Big[\frac{4}{\sqrt{3}\,\Delta}\Big(
f_{2}\,\psi_{2}(u,w)\,w\,(m_{b}+\bar u\,w) \nonumber\\
&-& f_{1}(\hat{\psi}_{3\sigma}(u,w)\,\bar u-\psi_{3s}(u,w)\,\bar u\,w^{2}\Big)
-\frac{4\,\bar u^{2}\,w}{\sqrt{3}\,\Delta^{2}}\Big(
2\,q\!\cdot\! v\,f_{1}\,\hat{\psi}_{3\sigma}(u,w)\nonumber\\
&+&f_{2}\big(\hat{\psi}_{2}(u,w)-\hat{\psi}_{4}(u,w))\,(m_{b}+\bar u\,w)\Big
)\Big]\,
q_{\mu}\,v_{\nu}\, +\cdots\Bigg\}u_{\Xi_{b}}(v)
.\label{Eq:QCDside}
\end{eqnarray}
The dots in Eq.~(\ref{Eq:QCDside}) represent the contributions coming from other structures and the functions
$\hat{\psi}(u,w)$ and $\Delta$ are given as
\begin{eqnarray}
\hat{\psi}(u,w)=\int_{0}^{w} d\tau\,\tau\,\psi(u,\tau)\,,
\label{eq:psihat_def}
\end{eqnarray}
and
\begin{eqnarray}
\Delta=\frac{p^{\prime 2}\,\bar{u}\,w}{m_{\Xi_b}}-m_{\Xi_b}\,\bar{u}\,w+\bar{u}^{2}w^{2}-m_{b}^{2}\,.
\label{eq:Delta_def}
\end{eqnarray}

Having completed the calculation of both sides of the correlation function, the next stage is equating the coefficients of the same structure in hadronic and QCD sides to obtain the sum rules for the form factors, $T_{1}^{V}$ , $T_{2}^{V}$, $T_{1}^{A}$ and $T_{2}^{A}$. The final step for obtaining the sum rules for the form factors is performing Borel transformation from both sides of the correlation function. Finally for the form factors we get the following sum rules:
\begin{eqnarray}
T_{1}^{V}
&=& e^{\frac{m_{\Xi^{*}_{bb}}^{2}}{M^{2}}}\, \frac{ m_{\Xi_{b}}\, \tilde{\Pi}_{2}^{\mathrm{QCD}} }{ \lambda }, \nonumber\\
T_{2}^{V}
&= &e^{\frac{m_{\Xi^{*}_{bb}}^{2}}{M^{2}}}\,\frac{ 
\big( -2\,m_{\Xi_{b}}\,\tilde{\Pi}_{1}^{\mathrm{QCD}}
      - m_{\Xi_{b}}^{2}\,\tilde{\Pi}_{2}^{\mathrm{QCD}}
      + m_{\Xi^{*}_{bb}}^{2}\,\tilde{\Pi}_{2}^{\mathrm{QCD}} \big)}
{ 2\,\lambda\,\big( m_{\Xi_{b}} + m_{\Xi^{*}_{bb}} \big) },\nonumber\\
T_{1}^{A}
&=&e^\frac{m_{\Xi^{*}_{bb}}^{2}}{M^{2}}\,  \frac{ m_{\Xi_{b}}\, \tilde{\Pi}_{3}^{\mathrm{QCD}} }{ \lambda }, \nonumber\\
T_{2}^{A}
&= & -e^\frac{m_{\Xi^{*}_{bb}}^{2}}{M^{2}}\,\frac{ 
\big( m_{\Xi_{b}}\,\tilde{\Pi}_{3}^{\mathrm{QCD}}
      + m_{\Xi^{*}_{bb}}\,\tilde{\Pi}_{3}^{\mathrm{QCD}}
      + \tilde{\Pi}_{4}^{\mathrm{QCD}} \big)}
{ 2\,\lambda } \, ,
\end{eqnarray}
where $\tilde{\Pi}_{i}^{\mathrm{QCD}}$ represents the Borel transformed results of the calculations obtained from the QCD side corresponding to the coefficients of the structures $\slashed{q}\gamma_5 g_{\mu\nu} $, $\slashed{q}\gamma_5 q_{\mu}v_{\nu}$, $\slashed{q} q_{\mu}v_{\nu}$ and $q_{\mu}v_{\nu}$ given in Eq.~(\ref{Eq:QCDside}). The Borel transformation and the continuum subtraction are performed with the help of following master formula: 
\begin{eqnarray}
\int_0^1 du \int dw\frac{\rho(u,w)}{\Delta^k}&=&(-1)^k\Bigg\{\int_0^1 du \int_0^{w_0}dwe^{-\frac{s}{M^2}}\frac{\rho(u,w)}{(k-1)!(\frac{\bar{u}w}{m_{\Xi_b}})(M^2)^{k-1}}\nonumber\\
&+&\int_0^1 du\Bigg[\frac{1}{(k-1)!}e^{-\frac{s_0}{M^2}}\sum_{j=1}^{k-1}\frac{1}{(M^2)^{k-j-1}}\frac{1}{s'}\Big(\frac{d}{dw}\frac{1}{s'}\Big)^{j-1}\frac{\rho(u,w)}{(\frac{\bar{u}w}{m_{\Xi_b}})^k}\Bigg]_{w=w_0}\Bigg\},
\label{Eq:MsterBorel}
\end{eqnarray}
where the integration region is the region in the $uw$ plane where $s > 4m _b^2$. In Eq.~(\ref{Eq:MsterBorel}) $s=m_{\Xi_b}\left(m_{\Xi_b}+\frac{m_{b}^{2}}{\overline{u}\,w}-\overline{u}\,w\right)$, and $w_0$ is obtained from the solution of the equation $s=s_{0}$, where $s_0$ is the continuum threshold.

The analytic results obtained in this section are used in the following section to obtain the results for the form factors numerically.

\section{Numerical Analyses}\label{III}

This section is devoted to numerical analyses of the analytic results for the form factors given in Sec.~\ref{sec:LCSR} at $q^2=0$. Additionally, using the obtained form factors we present the corresponding decay width.

In our numerical analysis the main input parameters are the DA's of the $\Xi_{b}$ baryon which are~\cite{Ali:2012zza}  
\begin{eqnarray}
\label{eq:DAsfunc}  
\psi_2(u,w) &=& w^2 \bar{u}u \sum_{n=0}^2 {a_n\over \varepsilon_n^4}
{C_n^{3/2} (2 u -1) \over | C_n^{3/2} |^2 } e^{-w/\varepsilon_n}~, \nnb \\
\psi_4(u,w) &=&  \sum_{n=0}^2 {a_n\over \varepsilon_n^2}
{C_n^{1/2} (2 u -1) \over | C_n^{1/2} |^2 } e^{-w/\varepsilon_n}~, \nnb \\
\psi_3^{(\sigma,s)}(u,w) &=& {w \over 2}  \sum_{n=0}^2 {a_n\over
\varepsilon_n^3}
{C_n^{1/2} (2 u -1) \over | C_n^{1/2} |^2 } e^{-w/\varepsilon_n}~.
\label{Eq:DAs}
\end{eqnarray}
The parameters $a_0,~a_1,~a_2$, and $\varepsilon_0,~\varepsilon_1,~\varepsilon_2$ in Eq.~(\ref{Eq:DAs}) are given in Ref.~\cite{Ali:2012pn}, $A=1/2$, $C_n^{\lambda}(2 u-1)$ is the Gegenbauer polynomial, and $|C_n^{\lambda}|^2$ is
\begin{eqnarray}
|C_n^{\lambda}|^2=\int_0^1 du [C_n^\lambda(2u-1)]^2.
\end{eqnarray}
Besides DA's, we also need other input parameters such as current coupling constant $\lambda=(0.22\pm 0.03)~\mathrm{GeV}^3$~\cite{Aliev:2012iv}, $f_1=f_2=(0.032\pm 0.009)~\mathrm{GeV}^3$~\cite{Wang:2010fq}, $|V_{tb}|=1.014\pm 0.029$, $|V_{ts}|=(41.5\pm 0.9)\times 10^{-3}$, $m_b = 4.78\pm 0.06$~GeV, $m_s=93.4^{+8.6}_{-3.4}$~MeV, $m_{\Xi{}_b^-}=(5797.0\pm 0.6)$~MeV~\cite{ParticleDataGroup:2024cfk}, $m_{\Xi^{*}{}_{bb}^-}=10.237$~GeV~\cite{Ebert:2002ig}.  

\begin{figure} []
\centering
\begin{tabular}{cccc}
\includegraphics[width=0.4\textwidth]{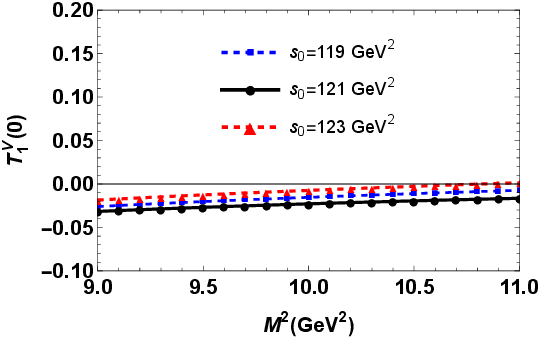} &
\includegraphics[width=0.4\textwidth]{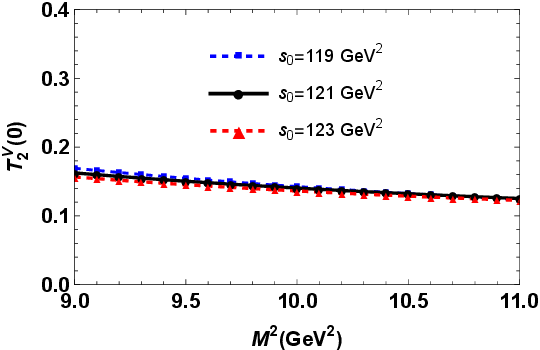} \\
\end{tabular}
\begin{tabular}{cccc}
\includegraphics[width=0.4\textwidth]{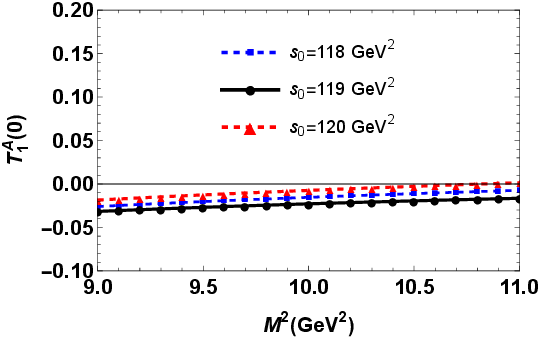} &
\includegraphics[width=0.4\textwidth]{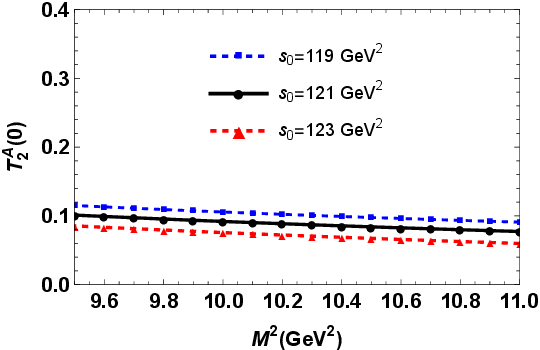} \\
\end{tabular}
\caption{ The variation of the form factors of $\Xi^{*}_{bb} \rightarrow \Xi_{b}\gamma$ transition as a function of $M^2$ at different values of $s_0$. 
}
\label{fig:1}
\end{figure}

In addition to these input parameters, in the results there are auxiliary parameters, such as Borel parameter $M^2$ and threshold parameter $s_0$.  The threshold parameter is related to the excited state of the considered baryon, and considering these its value is fixed to the interval $119~\mathrm{GeV}^2\leq s_0\leq 123~\mathrm{GeV}^2$. Imposing that the contribution of the leading twist-2 term dominates over the power corrections and continuum contributions, the interval for the Borel parameter is determined as $9~\mathrm{GeV}^2 \leq M^2 \leq 11~\mathrm{GeV}^2$. Fig.~\ref{fig:1} presents the dependence of the form factors , $T_{1}^{V}$ , $T_{2}^{V}$, $T_{1}^{A}$ and $T_{2}^{A}$, at $q^2=0$,  on Borel parameter $M^2$  within the given interval of the Borel parameter, and at different values of threshold parameters.  
To determine the numerical values of the form factors and their corresponding uncertainties, we carry out the analysis in which the Borel parameter $M^2$ and the threshold parameter $s_0$ are randomly sampled within their pre-specified intervals. With 5000 random sets of these parameters, we obtain the corresponding histograms presented in Fig.~\ref{fig:Hist}, and the mean values and standard deviations of the form factors. From these analyses we obtain the form factors as follows:
\begin{figure} []
\centering
\begin{tabular}{cccc}
\includegraphics[width=0.4\textwidth]{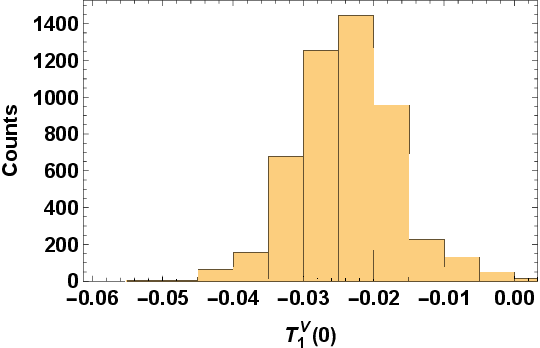} &
\includegraphics[width=0.4\textwidth]{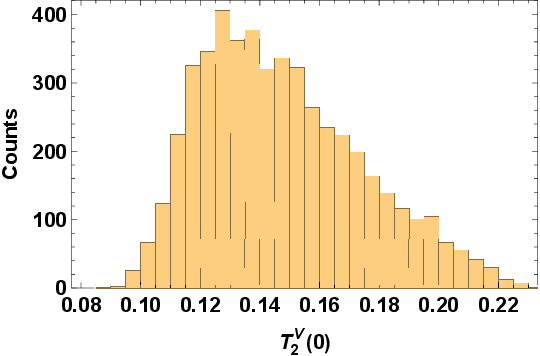} \\
\end{tabular}
\begin{tabular}{cccc}
\includegraphics[width=0.4\textwidth]{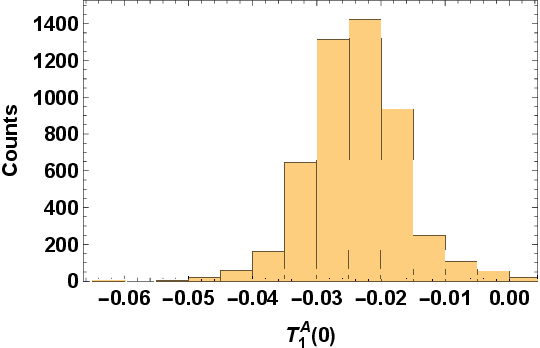} &
\includegraphics[width=0.4\textwidth]{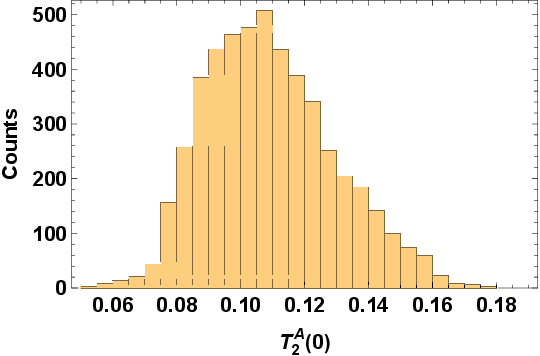} \\
\end{tabular}
\caption{ The histograms of the form factors obtained using randomly selected values of the auxiliary parameters $M^2$ and $s_0$.
}
\label{fig:Hist}
\end{figure}
\begin{eqnarray}
T_1^V(q^2=0) &=& -0.024\pm 0.007~\mathrm{GeV}, ~~~~~~~~~~T_2^V(q^2=0) = 0.15\pm 0.03~\mathrm{GeV} \nonumber\\
T_1^A(q^2=0) &=& -0.024\pm 0.007~\mathrm{GeV}, ~~~~~~~~~~
T_2^A(q^2=0)= 0.11\pm 0.02~\mathrm{GeV}. 
\label{Formfresults}
\end{eqnarray}
We see that the endpoint relations for $T_1^{V}$ and $T_1^{A}$, namely $T_1^{V}(0)=T_1^{A}(0)$, are satisfied exactly, and the relations between $T_2^{V}(q^{2})$ and $T_2^{A}(q^{2})$ are satisfied with good accuracy.

With the obtained values of the form factors given in Eq.~(\ref{Formfresults}), we predict the  width of considered decay. After some calculations for the decay width we get
\begin{eqnarray}
\Gamma &=&
\frac{\alpha_{\mathrm{em}}\,G_F^{2}\;
      \big|C_{7\gamma}^{(0)\,\mathrm{eff}}(m_b)\big|^{2}\;
      \big|V_{tb}V_{ts}^{*}\big|^{2}}{768\,\pi^{4}}\;
\frac{\big(m_{\Xi_{bb}^{*}}^{2}-m_{\Xi_b}^{2}\big)^{3}}{m_{\Xi_b}^{4}\,m_{\Xi_{bb}^{*}}^{5}}
\Bigg[
- 2\,m_b m_s\Big(
      m_{\Xi_{bb}^{*}}^{4}\big(|T_{1}^{A}|^{2}-|T_{1}^{V}|^{2}\big)
      \nonumber\\
      &+& m_{\Xi_b}\, m_{\Xi_{bb}^{*}}^{3}\big(2|T_{1}^{A}|^{2}+5\,\mathrm{Re}[T_{1}^{A}T_{2}^{A*}]
                              +2|T_{1}^{V}|^{2}-5\,\mathrm{Re}[T_{1}^{V}T_{2}^{V*}]\big)     
      + m_{\Xi_b}^{2} m_{\Xi_{bb}^{*}}^{2}\big(|T_{1}^{A}|^{2}-|T_{1}^{V}|^{2}
         \nonumber\\
& +& 4\,\mathrm{Re}[T_{1}^{A}T_{2}^{A*}]
         +7|T_{2}^{A}|^{2}
         +4\,\mathrm{Re}[T_{1}^{V}T_{2}^{V*}]
         -7|T_{2}^{V}|^{2}\big)
      + m_{\Xi_b}^{4}\big(|T_{2}^{A}|^{2}-|T_{2}^{V}|^{2}\big)
    \nonumber\\
&  - & m_{\Xi_b}^{3} m_{\Xi_{bb}^{*}}\big(\mathrm{Re}[T_{1}^{A}T_{2}^{A*}]
         +4|T_{2}^{A}|^{2}
         -\mathrm{Re}[T_{1}^{V}T_{2}^{V*}]
         +4|T_{2}^{V}|^{2}\big)\Big)+ m_b^{2}\Big(
      m_{\Xi_{bb}^{*}}^{4}\big(|T_{1}^{A}|^{2}+|T_{1}^{V}|^{2}\big)\nonumber\\[2pt]
&
      - &m_{\Xi_b}^{3} m_{\Xi_{bb}^{*}}\big(\mathrm{Re}[T_{1}^{A}T_{2}^{A*}]
         +4|T_{2}^{A}|^{2}
         +\mathrm{Re}[(T_{1}^{V}-4T_{2}^{V})T_{2}^{V*}]\big)
      + m_{\Xi_b}^{4}\big(|T_{2}^{A}|^{2}+|T_{2}^{V}|^{2}\big)
  \nonumber\\
&    +& m_{\Xi_b} m_{\Xi_{bb}^{*}}^{3}\big(2|T_{1}^{A}|^{2}
         +5\,\mathrm{Re}[T_{1}^{A}T_{2}^{A*}]
         -2|T_{1}^{V}|^{2}
         +5\,\mathrm{Re}[T_{1}^{V}T_{2}^{V*}]\big)
      + m_{\Xi_b}^{2} m_{\Xi_{bb}^{*}}^{2}\big(|T_{1}^{A}|^{2}
      \nonumber\\
&+&|T_{1}^{V}|^{2}
         +4\,\mathrm{Re}[T_{1}^{A}T_{2}^{A*}]
         -4\,\mathrm{Re}[T_{1}^{V}T_{2}^{V*}]
         +7(|T_{2}^{A}|^{2}+|T_{2}^{V}|^{2})\big)\Big)+ m_s^{2}\Big(
      m_{\Xi_{bb}^{*}}^{4}\big(|T_{1}^{A}|^{2}+|T_{1}^{V}|^{2}\big)\nonumber\\[2pt]
&
      -& m_{\Xi_b}^{3} m_{\Xi_{bb}^{*}}\big(\mathrm{Re}[T_{1}^{A}T_{2}^{A*}]
         +4|T_{2}^{A}|^{2}
         +\mathrm{Re}[(T_{1}^{V}-4T_{2}^{V})T_{2}^{V*}]\big)
      + m_{\Xi_b}^{4}\big(|T_{2}^{A}|^{2}+|T_{2}^{V}|^{2}\big) \nonumber\\
&
      +& m_{\Xi_b} m_{\Xi_{bb}^{*}}^{3}\big(2|T_{1}^{A}|^{2}
         +5\,\mathrm{Re}[T_{1}^{A}T_{2}^{A*}]
         -2|T_{1}^{V}|^{2}
         +5\,\mathrm{Re}[T_{1}^{V}T_{2}^{V*}]\big)
      + m_{\Xi_b}^{2} m_{\Xi_{bb}^{*}}^{2}\big(|T_{1}^{A}|^{2}+|T_{1}^{V}|^{2}
          \nonumber\\
&+& 4\,\mathrm{Re}[T_{1}^{A}T_{2}^{A*}]
         -4\,\mathrm{Re}[T_{1}^{V}T_{2}^{V*}]
         +7(|T_{2}^{A}|^{2}+|T_{2}^{V}|^{2})\big)\Big)
\Bigg].
\label{Eq:DW}
\end{eqnarray}
Using the numerical values of the quantities entering in Eq.~(\ref{Eq:DW}), for the decay with we obtain
\begin{equation}
\Gamma(\Xi^{*}_{bb}\rightarrow \Xi_{b}\gamma)=(1.23\pm 0.23)\times 10^{-19}~\mathrm{GeV}.
\end{equation}

At the end of this section we would like to note that the calculations are directly applicable for estimating
$\Xi_{bb}^{*}\to \Lambda_{b}\gamma$,
$\Xi_{cc}^{*}\to \Xi_{c}\gamma$,
$\Xi_{bc}^{*}\to \Xi_{b}(\Lambda_{b})\gamma$,
$\Xi_{cc}^{*}\to \Xi_{c}(\Lambda_{c})\gamma$
transitions as well. After making appropriate replacements one can immediately estimate the transition form factors as well as the corresponding decay widths. For instance considering the ratio,
\begin{equation}
\frac{\Gamma\!\left(\Xi_{bb}^{*}\to \Xi_{b}\gamma\right)}
{\Gamma\!\left(\Xi_{bb}^{*}\to \Lambda_{b}\gamma\right)}
\simeq
\frac{\left|V_{tb}V_{ts}^{*}\right|^{2}}{\left|V_{tb}V_{td}^{*}\right|^{2}}
\simeq 20\, ,
\end{equation}
in the SU(3) symmetry limit,
\begin{equation}
\Gamma\!\left(\Xi_{bb}^{*}\to \Lambda_{b}\gamma\right)
\simeq \frac{1}{20}\,\Gamma\!\left(\Xi_{bb}^{*}\to \Xi_{b}\gamma\right)
\simeq 6\times 10^{-21}\ \mathrm{GeV}\, .
\end{equation}

\section{Summary and conclusion}\label{IV}

In the present work, the form factors of the rare radiative $\Xi^{*}_{bb}\rightarrow\Xi_{b} \gamma$ decay were calculated using the light cone QCD sum rules method and the heavy $\Xi_b$ baryon distribution amplitudes. At the quark level, the process is driven by the flavor changing neutral current transition $b\to s\gamma$, and the decay amplitude can be written in terms of four tensor form factors $T_{1}^{V}(q^2)$, $T_2^{V}(q^2)$, $T_{1}^{A}(q^2)$, $T_2^{A}(q^2)$. The numerical values of the form factors were obtained at $q^2=0$ as $T_{1}^{V}(0) = -0.024\pm0.007$~GeV, $T_2^V(0) = 0.15\pm0.03$~GeV, $T_1^A(0) = -0.024\pm0.007$~GeV, and $T_2^A(0) = 0.11\pm0.02$~GeV, which satisfy the endpoint relation within their uncertainties. Using the obtained form factors, we computed the decay width as $\Gamma(\Xi^{*}_{bb}\rightarrow \Xi_{b}\gamma)=(1.23\pm 0.23)\times 10^{-19}~\mathrm{GeV}$. The presented calculation scheme is applicable for estimating corresponding rare radiative decays, like
$\Xi_{cc}^{*}\to \Xi_{c}(\Lambda_{c})\,\gamma$,
$\Xi_{bc}^{*}\to \Xi_{b}(\Lambda_{b})\,\gamma$,
$\Xi_{bc}^{*}\to \Xi_{c}(\Lambda_{c})\,\gamma$ as well. Although the predicted width is small, it falls in a range that may become accessible with future high luminosity data on doubly heavy baryons at LHCb or similar experiments.


\bibliographystyle{apsrev4-1}  
\bibliography{refs}


\end{document}